\documentclass[aps,prb,reprint,amssymb,amsmath,superscriptaddress]{revtex4-1}

\usepackage{graphicx}

\begin{document}

\title{Interaction correction to the conductivity of two-dimensional electron
gas in In$_x$Ga$_{1-x}$As/InP quantum well structure with strong
spin-orbit coupling }

\author{G.~M.~Minkov}
\affiliation{Institute of Metal Physics RAS, 620990 Ekaterinburg,
Russia}

\affiliation{Institute of Natural Sciences, Ural Federal University,
620000 Ekaterinburg, Russia}

\author{A.~V.~Germanenko}

\author{O.~E.~Rut}
\affiliation{Institute of Natural Sciences, Ural Federal University,
620000 Ekaterinburg, Russia}

\author{A.~A.~Sherstobitov}
\affiliation{Institute of Metal Physics RAS, 620990 Ekaterinburg,
Russia}

\affiliation{Institute of Natural Sciences, Ural Federal University,
620000 Ekaterinburg, Russia}

\date{\today}

\begin{abstract}

The electron-electron interaction quantum correction to the
conductivity of the gated single quantum  well
InP/In$_{0.53}$Ga$_{0.47}$As heterostructures is investigated
experimentally. The analysis of the temperature and magnetic field
dependences of the conductivity tensor allows us to obtain reliably the
diffusion part of the interaction correction for different values of
spin relaxation rate, $1/\tau_s$. The surprising result is that the
spin relaxation processes do not suppress the interaction correction in
the triplet channel and, thus, do not enhance the correction in
magnitude contrary to theoretical expectations even in the case of
relatively fast spin relaxation, $1/T\tau_s\simeq (20-25)\gg 1$.

\end{abstract} \pacs{73.20.Fz, 73.61.Ey}

\maketitle

\section{Introduction}
\label{sec:intr}

A weak localization (WL) quantum correction  to the conductivity,
$\delta\sigma^{WL}$, and the correction due to electron-electron ({\it
e-e}) interaction, $\delta\sigma^{ee}$, are responsible for the
temperature and magnetic field dependences of the conductivity and Hall
effect in the two-dimensional electron gas at low temperatures, when
the Fermi energy ($E_F$) is much greater than the temperature ($T$) and
the phonon scattering is negligible. Magnitude of these corrections is
strongly dependent on the spin-relaxation rate $1/\tau_s$, which is
determined by strength of spin-orbit interaction (SOI). When the spin
relaxation is slow, $\tau_s\gg\tau_\phi$, as it takes place in the
systems with the weak spin-orbit coupling, the WL correction is
negative and for the case of the one valley energy spectrum is
$\delta\sigma^{WL}=-G_0\,\ln(\tau_\phi/\tau)$, where $\tau_\phi$ and
$\tau$ are the phase relaxation and transport times, respectively,
$G_0=e^2/\pi h$. Applying a transverse magnetic field destroys the
interference resulting in negative magnetoresistance. Namely analysis
of the shape of the negative magnetoresistance is a powerful tool
allowing us to extract experimentally the phase relaxation time in
different systems.\cite{Hik80,Chak86,Wit87} The increase of SOI
strength leads to shortage of the spin relaxation time and to
appearance of the positive magnetoresistivity against the negative
magnetoresistivity background when $\tau_s\lesssim\tau_\phi$. This
effect known as the weak antilocalization has been first reported for
2D structures about two decades ago\cite{Dres92,Chen93} and widely used
to date to obtain the spin relaxation time
experimentally.\cite{Knap96,Knap96-1} In the limiting case of
$\tau_s\ll \tau_\phi$, the temperature dependence of the interference
correction is metallic-like,\cite{Lyanda98}
$\delta\sigma^{WL}=0.5\,G_0\,\ln(\tau_\phi/\tau)$, and the
magnetoconductivity becomes positive over the whole magnetic field
range.

The spin relaxation influences the interaction correction as
well.\cite{AA85} This correction can be conventionally considered as
the sum of two parts. The first part originates from the exchange
contribution, and the second one from the Hartree contribution. They
are usually referred as the correction in the singlet and triplet
channels. The singlet exchange conductivity correction is independent
of the magnetic field and  spin relaxation rate. It favors the
localization and is equal to $G_0\ln(T\tau)$ in the diffusion regime,
$T\tau\ll 1$. The triplet Hartree conductivity correction is
antilocalizing and it depends on the {\it e-e} interaction constant.
When the {\it e-e} interaction is not very strong, $k_F r_0\gtrsim
0.1$, where $k_F$ and $r_0$ are the Fermi quasimomentum and screening
length, respectively, the triplet contribution is comparable in
magnitude with the singlet one, although its absolute value is somewhat
less, and the total interaction correction is localizing.\cite{Zala01}
In contrast to the singlet contribution, the triplet term is sensitive
to the external magnetic field and spin relaxation processes. The
magnetic field suppresses two of three triplet contributions
\cite{Cast84-2,Fin84,Raim90,Cast98,Zala02,Gor04,Min05-1} due to the
Zeeman effect. As a result the total interaction correction consists of
the singlet and one triplet contributions when the magnetic field is
rather high, $\textsl{g}\mu_B B\gg T$, where $\textsl{g}$ is the
effective Land\'{e} $\textsl{g}$-factor. The spin relaxation processes
suppresses all the three triplet contributions already at $B=0$ if
$T\tau_s\ll 1$.\cite{Altshuler83,Lyanda98} In this case the whole
interaction correction comprises only the singlet term and should not
depend on the magnetic field.

As far as we know, the suppression of the triplet contributions in
systems with the growing spin relaxation rate predicted theoretically
did not observed experimentally, although it is invoked to interpret
the experimental results in different 2D
systems.\cite{Gornyi98,Skvortsov98} This is not surprising, because a
reliable extraction of the {\it e-e} interaction contribution to the
conductivity is a intricate challenge. To solve it, one needs, as a
rule, to investigate special structures with the given disorder
strength, conductivity and $T\tau$ values.\cite{Min03-2} A clear
indication of the complexity of this problem is the fact that the
quantitative determination of the interaction contributions and the
elucidation of the role of intervalley transitions is a widely debated
topic even in the popular Si based MOS structures.

In this paper we present the results of the experimental studies of the
{\it e-e} interaction contribution to the conductivity of the single
quantum well heterostructures InP/In$_x$Ga$_{1-x}$As/InP, $x=0.53$.
Earlier, the WL and {\it e-e} interaction corrections in analogous
structures but with lower indium content, $x\simeq 0.2$, are thoroughly
studied in a number of papers (see  Refs.
\onlinecite{Minkov06,Minkov07-1,Minkov09} and references therein). The
spin-orbit interaction in such type of structures is rather weak and
the experimental results are consistent with the well known
theories.\cite{AA85,Chak86,Wit87,Cast84-2,Fin84,Raim90,Cast98} The
increase of the indium content results in the narrowing of the energy
gap, decrease of the electron effective mass, increase of the effective
$\textsl{g}$ factor and, what is more important, to the strengthening
spin-orbit interaction. The last effect leads to reduction in the spin
relaxation time so that the condition  $ 1/T\tau_s\gg 1$ can be
fulfilled at easily achievable temperatures, $T\gtrsim 1$~K. The WL
effects in In$_x$Ga$_{1-x}$As quantum wells with relatively high indium
content, $x\simeq 0.53$, is thoroughly investigated in
Refs.~\onlinecite{Studenikin03,Yu08}. The authors show that the
theory\cite{Zdu97} developed within the framework of the diffusion
approximation is unable to describe the data for the high mobility
structures. To interpret the data, the authors attract the
theory\cite{Golub05,Glazov06} working over the whole ranges of the
temperature and classically weak magnetic fields. Thus, the
In$_x$Ga$_{1-x}$As, quantum wells with $x\simeq0.5$ are suitable
systems to study the effect of the spin-orbit interaction on the {\it
e-e} interaction contribution to the conductivity. Analyzing the
temperature and magnetic filed dependences of the conductivity
components over the temperature range from $0.6$~K to $4.2$~K for the
electron densities $(1.3 - 3.0)\times 10^{11}$~cm$^{-2}$ we find no
suppression of the interaction correction in the triplet channel up to
$1/T\tau_s\simeq 25$  contrary to the theoretical
expectations.\cite{Altshuler83,AA85,Lyanda98}

\section{Theoretical background}
\label{sec:theor}

In the absence of the magnetic field, the interaction correction in the
diffusion regime, $T\tau \ll 1$, is given by
\cite{AA85,Finkelstein83,Finkelstein84,Cast84-1,Cast84-2,Cast98}
\begin{equation}
 \frac{\delta\sigma^{ee}}{G_0}=K_{ee}\ln{T\tau},
  \label{eq00}
\end{equation}
where
\begin{equation}
 K_{ee}=1+3\left[1-\frac{1+\gamma_2}{\gamma_2}\ln\left(1+\gamma_2\right)\right]
   \label{eq10}
\end{equation}
with $\gamma_2$ standing for the Landau's Fermi liquid amplitude. The
two terms in Eq.~(\ref{eq10}) are the singlet and triplet channels
mentioned above. The specific feature of the interaction correction in
the diffusion regime is the fact that in the presence of magnetic field
it contributes to $\sigma_{xx}$ and not to $\sigma_{xy}$.\cite{AA85} As
already noted, the magnetic field does not change the singlet
contribution but suppresses the triplet one so that only one of three
triplet parts is alive when the Zeeman splitting $\textsl{g}\mu_BB$
becomes much larger than the temperature. The magnetic field dependence
of $\delta\sigma^{ee}_{xx}$ is given by
\begin{eqnarray}
 \frac{\delta\sigma^{ee}_{xx}}{G_0}&=&\left[1+1-\frac{1+\gamma_2}
 {\gamma_2}\ln\left(1+\gamma_2\right)
 \right]\ln{T\tau} \nonumber \\
 &+&2\left[1-\frac{1+\gamma_2}
 {\gamma_2}\ln\left(1+\gamma_2\right)
 \right]\ln{F(B,T\tau)},
  \label{eq20}
\end{eqnarray}
where $F(B,T\tau)$ is a function describing the suppression of the two
triplet channels.\cite{Finkelstein83,Cast98} Since  $F(B,T\tau)$
obtained in these papers is rather complicated, it is more convenient
to use in practice the following simple expression\cite{Min05-1}
\begin{eqnarray}
F(B,T\tau)=T\tau\sqrt{1+\left(\frac{\textsl{g}\mu_BB}{T}\right)^2},
\label{eq30}
\end{eqnarray}
which well approximates the results from
Refs.~\onlinecite{Finkelstein83,Cast98}. In contrast to the Zeeman
effect, the spin-orbit interaction should suppress all the three
triplet contributions so that the interaction correction consists of
only one singlet channel when $1/T\tau_s\gg 1$ even in zero magnetic
field. To the best of our knowledge, it is unknown until the present
how the triplet contribution is functionally suppressed with the
growing spin relaxation rate. Intuitively it can be written by analogy
with Eqs.~(\ref{eq20}) and (\ref{eq30}) as follows
\begin{eqnarray}
{\delta \sigma^{ee}_{xx}\over G_0}&=&\ln{T\tau}+
3\left[1-\frac{1+\gamma_2}{\gamma_2}\ln\left(1+\gamma_2\right)\right]
\nonumber \\
 &\times & \ln{T\tau\sqrt{1+(T\tau_s)^{-2}}}
\label{eq40}.
\end{eqnarray}
If one uses the parameters which are typical for the samples
investigated ($\gamma_2=0.6$, $\tau_s=2\times 10^{-12}$~s and
$\tau=3\times 10^{-13}$~s), we obtain that the triplet contribution
[the second term in Eq.~(\ref{eq40})] to the temperature dependence of
$\delta \sigma^{ee}_{xx}$ becomes negligibly small as compared with the
singlet one (the first term) at $1/T\tau_s\simeq 4$ corresponding to
$T=1$~K.

\section{Experimental details}
\label{sec:exp} We studied a InP/In$_x$Ga$_{1-x}$As/InP quantum well
structure grown by chemical beam epitaxy on an InP (100)
substrate.\cite{Lefebvre02} The structure CBE06-173 consists of a
$50$-nm Be-doped layer followed by $75$-nm undoped InP buffer layer,
$12.5$-nm Si-doped InP layer, a $12.5$~nm spacer of undoped InP, a
$10$~nm In$_{0.53}$Ga$_{0.47}$As quantum well, and $37.5$~nm cap layer
of undoped InP. The samples were etched into standard Hall bars. To
change the electron density in the well, an Al gate electrode was
deposited by thermal evaporation onto the cap layer through a mask. In
some cases the electron density and the conductivity were controlled
through  the illumination due to the persistent photoconductivity
effect. The results for equal electron densities were mostly identical
in both cases. Experiments were performed in a He$^3$ system with
temperatures from $4.2$~K down to $0.5$~K.

\section{Results and discussion}
\label{sec:res}

\begin{figure}[t]
\includegraphics[width=\linewidth,clip=true]{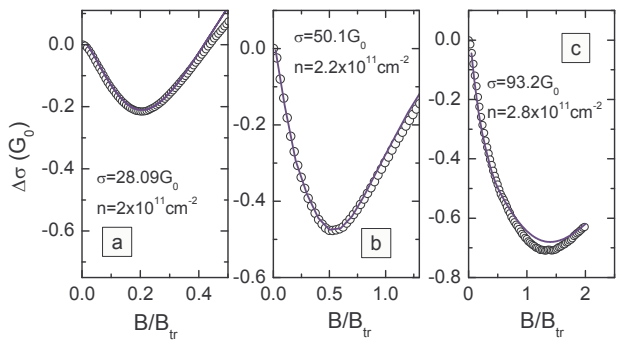}
\caption{(Color online) The magnetic field dependences of $\Delta\sigma$ for the different
conductivity controlled by the gate voltage at $T=1.35$~K. Symbols are the data,
the curves are the results of the best fit by the theory.\cite{Golub05,Glazov06}  }\label{f1}
\end{figure}

Because the leading parameter determining the suppression of the
interaction correction in the triplet channel is $T\tau_s$ value, we
begin our consideration with the determination of the spin relaxation
time in the sample investigated. One of the way to do this is the
analysis of the low magnetic field magnetoconductivity
$\Delta\sigma(B)=1/\rho_{xx}(B)-1/\rho_{xx}(0)$ caused by the
suppression of the interference quantum correction. Figure \ref{f1}
shows the low-field magnetoconductivity measured at $T=1.35$~K as a
function of a normalized magnetic field $B/B_{tr}$ for different
conductivity values, where $B_{tr}=\hbar/2el^2$ with $l$ as the mean
free path is the transport magnetic field. One can see that the
negative magnetoconductivity in low magnetic fields followed by the
positive magnetoconductivity in higher ones is observed for all the
cases. Such the behavior is the clear indication of the fact that the
spin relaxation time is shorter than $\tau_\phi$. The $\tau_\phi$ and
$\tau_s$ values are obtained from the fitting of the experimental data
by the theoretical expression. It should be emphasized that the
reliable determination of these values is possible only in the case
when the fit is done in the magnetic fields involving the region in
which the antilocalization minimum is observed. The expression obtained
in Refs.~\onlinecite{Knap96,Knap96-1} for the diffusion regime, when
$\tau/\tau_\phi,\,\tau/\tau_s,\,B/B_{tr}\ll 1$, is widely used for this
purpose. However it is inapplicable for our case, because the
antilocalization minimum is observed at relatively high magnetic
fields, $B\sim B_{tr}$, as seen from Figs.~\ref{f1}(b) and \ref{f1}(c).
As shown in Ref.~\onlinecite{Yu08} the phase and spin relaxation times
can be reliably obtained in these samples by fitting the experimental
data using the model,\cite{Golub05,Glazov06} which is developed for
arbitrary values of $\tau$, $\tau_\phi$, $\tau_s$, and magnetic field.
Inspection of Fig.~\ref{f1} shows that the
theory\cite{Golub05,Glazov06} describes the experimental results rather
well that allows us to extract the spin relaxation  and phase breaking
times for the different temperatures and the conductivity values.

Note the temperature dependences  of $\tau_\phi$ and $\tau_s$ (not
shown in the figures) are reasonable over the whole conductivity range,
$\sigma\simeq (20-100)\,G_0$. The spin relaxation time is independent
of the temperature within the accuracy of the experiment. Such the
behavior is natural for the degenerate electron gas ($E_F/T>20$ under
our experimental conditions). The temperature dependence of $\tau_\phi$
is close to $1/T$ that is typical for the 2D electron gas at low
temperatures when the main dephasing mechanism is the inelasticity of
the {\it e-e} interaction.\cite{AA85}

\begin{figure}
\includegraphics[width=0.75\linewidth,clip=true]{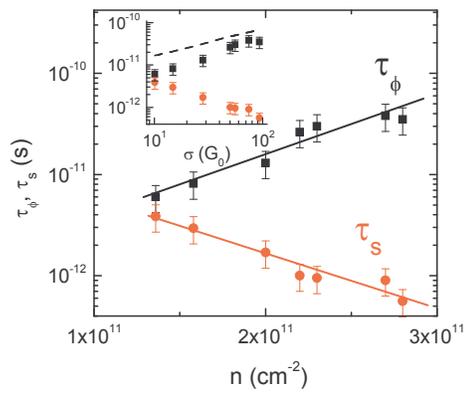}
\caption{(Color online) The phase and spin relaxation times at $T=1.35$~K
as a function of electron density and conductivity (in the inset).
Symbols are the data, the solid lines are provided
as a guide for the eye. The dashed line is calculated according to Ref.~\onlinecite{Nar02}.}\label{f2}
\end{figure}

Figure~\ref{f2} shows the spin relaxation and phase breaking time
plotted against the electron density and conductivity. One can see that
$\tau_s$ is shorter than $\tau_\phi$ for all the densities. The
dephasing time increases with the increasing conductivity that agrees
with the theoretical prediction.\cite{Altshuler82,Nar02} The spin
relaxation time decreases with the $n$ increase. The origin is
transparent. As shown in Refs.~\onlinecite{Studenikin03,Yu08} the spin
relaxation in such the type of samples at low temperatures is
controlled by the Dyakonov-Perel mechanism.\cite{Dyakonov71e} For this
mechanism the spin relaxation time is determined both by the spin orbit
splitting  at the Fermi energy $\hbar\Omega$ and the transport
relaxation time: $\tau_s=1/(2\Omega^2\tau)$. As shown in
Ref.~\onlinecite{Yu08} the spin orbit splitting in these samples is
caused by the Rashba effect\cite{Rash84} and it linearly depends on the
Fermi quasimomentum $k_F$: $\hbar\Omega=\alpha k_F=\alpha\sqrt{2\pi n}$
where $\alpha$ is the constant depending on the asymmetry of the
quantum well. So, since both quantities $\tau$ and $\hbar\Omega$
increase when $n$ increases, the decrease of $\tau_s$ in Fig.~\ref{f2}
observed experimentally  is natural.

Thus we have obtained the phase breaking and spin relaxation times over
the whole conductivity range. As seen from Fig.~\ref{f2} the spin
relaxation time decreases from  $\tau_s\approx4\times 10^{-12}$~s to
$\tau_s\approx5\times 10^{-13}$~s with changing electron density within
the range from $1.4\times 10^{11}$~cm$^{-2}$  to $2.8\times
10^{11}$~cm$^{-2}$. It means that parameter $\hbar/\tau_s=(2-15)$~K is
greater or much greater than the temperature under our experimental
conditions and the sample is good candidate for studying the role of
spin effects in the \emph{e-e} interaction correction to the
conductivity.

\begin{figure}[t]
\includegraphics[width=\linewidth,clip=true]{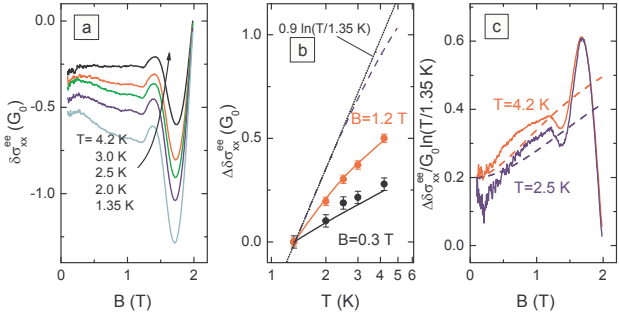}
\caption{(Color online) (a) -- The experimental magnetic field dependences
of $\delta\sigma_{xx}^{ee}$ for the different temperatures. (b) -- The dependences
$\Delta\delta\sigma_{xx}^{ee}(T)=\delta\sigma_{xx}^{ee}(T)-\delta\sigma_{xx}^{ee}(1.35\text{ K})$ for $B=0.3$~T and
$1.2$~T. The symbols are the data, the solid lines
are calculated from Eqs.~(\ref{eq20}) and (\ref{eq30}), the dashed line is calculated from Eq.~(\ref{eq40}). (c) -- The magnetic
field dependences of $\Delta\delta\sigma_{xx}^{ee}(T)/\ln{(T/1.35~\text{K})}$ for $T=4.2$~K
and $2.5$~K. Solid lines are the experimental results. Dashed lines are calculated
from Eq.~(\ref{eq40}).
The results presented in all the panels are obtained at $n= 2\times 10^{11}$~cm$^{-2}$,
$\sigma(T=1.35~\text{Ê})=28.09\,G_0$. }\label{f3}
\end{figure}

Now we are in position to consider the interaction quantum correction
to the conductivity. The diffusion contribution
$\delta\sigma_{xx}^{ee}$ can be obtained by eliminating the
interference correction and the ballistic part of interaction
correction with the use of the method described in
Ref.~\onlinecite{Min03-2}. Because both the interference correction and
the ballistic part are reduced to the renormalization of the mobility
and, therewith, the diffusion part of the correction does not
contribute to the off-diagonal component of the conductivity, one can
obtain the mobility $\mu$ from $\sigma_{xy}$ knowing the electron
density (from the period Shubnikov-de Haas oscillations)
$\mu(B,T)=\sqrt{\sigma_{xy}/(en-\sigma_{xy}B)B}$ and find the
correction $\delta\sigma_{xx}^{ee}$ as the difference between the
experimental value of $\sigma_{xx}$ and the value of
$en\mu/(1+\mu^2B^2)$. As an example we demonstrate the $B$ dependences
of $\delta\sigma_{xx}^{ee}$ obtained in such a way at different
temperatures for $n= 2\times 10^{11}$~cm$^{-2}$  in Fig.~\ref{f3}(a).
Two specific features are evident.  The lower temperature the more
negative the interaction correction $\delta\sigma_{xx}^{ee}$. The
$\delta\sigma_{xx}^{ee}$ value increases in magnitude with the growing
magnetic field.

The experimental temperature dependences of
$\Delta\delta\sigma_{xx}^{ee}$ taken at different magnetic fields are
shown by symbols in Fig.~\ref{f3}(b). One can see that the dependence
$\Delta\delta\sigma_{xx}^{ee}(T)=\delta\sigma_{xx}^{ee}(T)-\delta\sigma_{xx}^{ee}(1.35\text{
K})$ is close to the logarithmic one at  low magnetic field, $B=0.3$~T,
and perceptibly deviates from that at stronger field, $B=1.2$~T.
Furthermore, it  becomes steeper with growing  magnetic field that is
better evident from Fig.~\ref{f3}(c) where the quantity
$\Delta\delta\sigma_{xx}^{ee}(T)/\ln{(T/1.35~\text{K})}$ for $T=4.2$~K
and $2.5$~K against the magnetic filed is depicted.

Such the behavior of $\delta\sigma_{xx}^{ee}$ with growing temperature
and magnetic field is typical for the systems the relatively large
value of $\textsl{g}$-factor (see, for instance,
Ref.~\onlinecite{Minkov07}).  In the low magnetic field, the Zeeman
splitting is small as compared with the temperature. Really, if one
uses $\textsl{g}=2$ determined experimentally from the analysis of the
angle dependence of the Shubnikov-de Haas oscillations amplitude, we
obtain $\textsl{g}\mu_BB/T\simeq 0.2$ at $T=2$~K and  $B= 0.3$~T. In
this case $F(B,T\tau)\simeq T\tau$ as it follows from Eq.~(\ref{eq30}),
i.e., the triplet contribution is not suppressed and the temperature
dependence of $\delta\sigma_{xx}^{ee}$ is logarithmic with the slope
determined only by the value of interaction constant $\gamma_2$
according to Eq.~(\ref{eq10}). Because the Zeeman splitting increases
with the growing magnetic field, its role becomes more significant,
that results in suppression of the triplet contribution. This effect
manifests itself as the increase of absolute value of
$\delta\sigma_{xx}^{ee}$ and the increase of the slope of the
$\Delta\delta\sigma_{xx}^{ee}$~vs~$\ln{T}$ dependence with increasing
magnetic field. As seen from Figs.~\ref{f3}(b) and \ref{f3}(c) the
experimental data are well described in framework of this model. The
curves calculated from  Eq.~(\ref{eq30}) with $\gamma_2=0.64$ [that
corresponds to $K_{ee}=0.2$ in Eq.~(\ref{eq10})] and $\textsl{g}=2$ run
practically over the experimental points.

Thus we do not detect the suppression of the {\it e-e} interaction
correction by the spin-orbit interaction at $B=0$ when investigating
the magnetoresistivity in the transverse magnetic field. The
experimental results are well described by the theory taking into
account only the Zeeman effect, even though the triplet contribution
and, hence, the magnetic field dependence of the whole interaction
correction are expected to be strongly suppressed  due to the fast spin
relaxation.

\begin{figure}[t]
\includegraphics[width=\linewidth,clip=true]{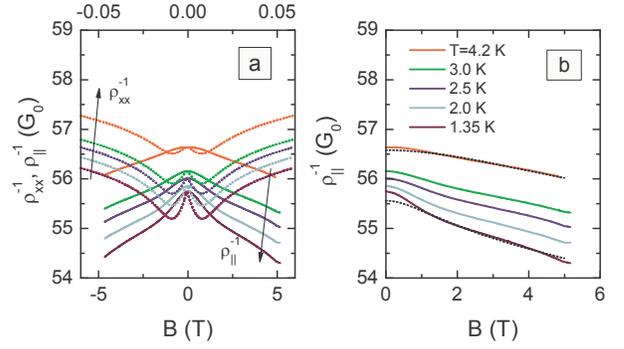}
\caption{(Color online) (a) -- The experimental magnetic field dependences
of the conductivity measured in the transverse (dotted curves) and longitudinal (solid curves)
orientations of the magnetic field for the different temperatures. (b) -- The longitudinal
magnetoconductivity for different temperatures measured experimentally (solid curves) and calculated
(dotted curves) according to Ref.~\onlinecite{Zala02}. }\label{f4}
\end{figure}

It would seem that more straightforward way to investigate the {\it
e-e} interaction correction is to take the measurements in in-plane
magnetic field,\cite{Zala02} where the off-diagonal component of the
conductivity tensor is equal to zero. The experimental magnetic field
dependences of the conductivity measured in the transverse (dotted
curves) and longitudinal (solid curves) orientations of the magnetic
field measured at the different temperatures for $n=2.2\times
10^{11}$~cm$^{-2}$ are shown in Fig.~\ref{f4}(a). The nonmonotonic
behavior of the  transverse magnetoconductivity is caused by the
suppression of the interference quantum correction in this magnetic
field range. As for the longitudinal orientation,  one can see that two
regions of the magnetic field can be distinguished (note, the scale for
the longitudinal orientation is hundred times larger than that for the
transverse one). This is region of low field, $B\lesssim 1.5$~T,
corresponding to the faster decrease of the conductivity
$\rho^{-1}_{\parallel}$ with growing magnetic field, and the region of
the higher magnetic filed characterized by the smaller curvature of the
magnetoconductivity curves. The behavior of $\rho^{-1}_{\parallel}$ in
the first region, in which the conductivity decreases on the value
close to the depth of the antilocalization minimum observed in
transverse magnetic field, is dictated by the suppression of the weak
antilocalization caused by roughness of the quantum well
interfaces.\cite{Mens87,Math01,Min04} The run of the
magnetoconductivity curves in higher magnetic field is controlled by
the suppression of the triplet contribution to the interaction-induced
quantum correction due to the Zeeman effect. The latter is clearly
evident from Fig.~\ref{f4}(b), where our data are presented together
with the results of the theoretical calculations.\cite{Zala02} The
theoretical curves were calculated with $F_0^\sigma=-0.39$ for the
diffusion part of the correction (that corresponds to
$\gamma_2=-F_0^\sigma/(1+F_0^\sigma)=0.64$ found above), and
$F_0^\sigma=-0.07$ for the ballistic part.

\begin{figure}
\includegraphics[width=0.75\linewidth,clip=true]{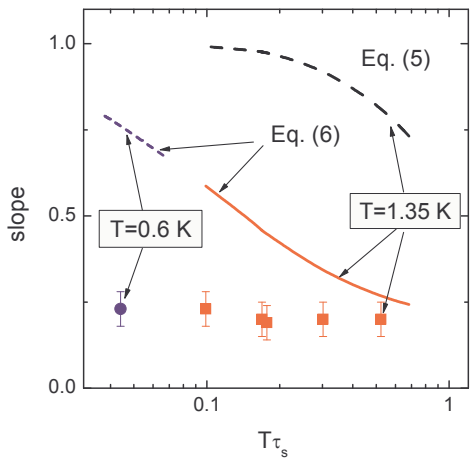}
\caption{(Color online) The slope of the temperature dependence of the {\it e-e} interaction
correction, $G_0^{-1}d\sigma/d{\ln{T\tau}}$, as a function of $T\tau_s$, calculated from
Eq.~(\ref{eq40}) and Eq.~(\ref{eq50}) (lines) and obtained experimentally (symbols)
for $T=0.6$~K and $1.35$~K.  }\label{f5}
\end{figure}

It must be noted here that the experiments in in-plane magnetic filed
cannot independently provide a reliable information about the singlet
and triplet contributions to the interaction correction. The reason is
that the two different interaction constants are responsible for the
ballistic and diffusion contributions to interaction
correction,\cite{Zala01} and the data for this orientation can be
satisfactorily described by different sets of these constants.
Nevertheless, we would like to stress that the results for the in-plane
orientation of magnetic field are consistent with that obtained in the
transverse orientation: the longitudinal magnetoconductivity is well
described by the interaction constant, which value, $F_0^\sigma=-0.39$,
has been obtained from the analysis of the $T$ and $B$ dependences of
$\sigma_{xx}$.

Thus, the experimental results obtained for both orientations  of  the
magnetic  field  are  adequately described by taking into account only
the Zeeman effect resulting in suppression of the two of three triplet
contributions with growing magnetic field.

This observation is in conflict with our expectations propounded in
Section~\ref{sec:theor}. It is well evident from Fig.~\ref{f5}, where
the data for different electron density are collected. Squares in this
figure are the experimental values of $K_{ee}$ at $B=0$ plotted against
the value of $T\tau_s$, where $\tau_s$ is controlled by the electron
density, while $T$ is kept constant, $T=1.35$~K. The dashed line is the
slope of the $T$ dependence of the interaction correction,
$G_0^{-1}d\sigma^{ee}/d\ln(T\tau)$, calculated from Eq.~(\ref{eq40}).
According to Eq.~(\ref{eq40}) the triplet channel should be fully
suppressed by SOI already at $T\tau_s\simeq 0.3$ (it corresponds to
$\tau_s=2\times 10^{-12}$~s for $T=1.35$~K); the slope is expected to
be equal to approximately $0.9$ instead of $0.2$ observed
experimentally.

A more accurate expression for the interaction correction in the
presence of spin-orbit interaction (assuming the simplest case when all
the three triplets are suppressed by the same $\tau_s$) can be written
\cite{PrvGornyi} similarly to the expression for the Zeeman
magnetoresistance:\cite{Zala02,Gor04}
\begin{eqnarray}
{\delta \sigma^{ee}\over G_0}&=&\ln{x}-
3\sum_{n=1}^{1/x} \left[\frac{y}{ny+1}\right.
\nonumber \\
 &+&\left.\frac{1+\gamma_2}{\gamma_2}\frac{1}{n}\ln{\frac{ny+1}{(1+\gamma_2)ny+1}}\right],\,\,\,x\ll 1,
\label{eq50}
\end{eqnarray}
where $x=T\tau$ and $y=2\pi T\tau_s$. The suppression of the triplet
contribution with growing spin relaxation rate occurs not so fast
according to this equation (see solid line in Fig.~\ref{f5}).
Nevertheless, the discrepancy between the experimental results and more
refined theoretical prediction remains well visible, especially at low
$T\tau_s$ values.

\begin{figure}[t]
\includegraphics[width=0.75\linewidth,clip=true]{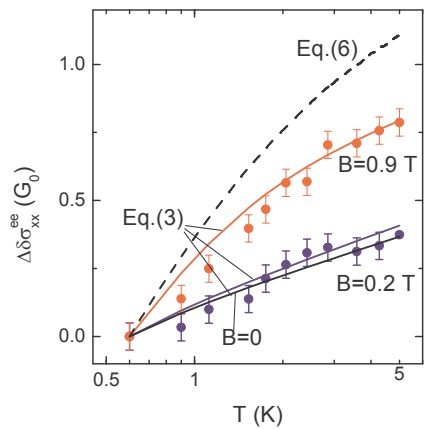}
\caption{(Color online) The temperature dependence of $\Delta\delta\sigma^{ee}_{xx}=\delta\sigma^{ee}_{xx}(T)-
\delta\sigma^{ee}_{xx}(0.6\text{ K})$ for $n=2.8\times10^{11}$~cm$^{-2}$ at different
magnetic fields. The symbols are the experimental results.  The solid and dashed lines are calculated
according to Eq.~(\ref{eq20}) and Eq.~(\ref{eq50}), respectively. The following parameters have been used:
$\gamma_2=0.653$, $\textsl{g}=2$, $\tau=6.5\times 10^{-13}$~s, and $\tau_s=6\times 10^{-13}$~s.}\label{f6}
\end{figure}

The contradiction between the data and theory becomes all the more
evident at lowest temperature, $T=0.6$~K, that provides $T\tau_s\simeq
0.05$ for the highest electron density, $n=2.8\times10^{11}$~cm$^{-2}$,
when $\tau_s=6\times 10^{-13}$~s). The results are shown in
Fig.~\ref{f6}. One can see that the temperature dependences of
$\delta\sigma_{xx}^{ee}$ in this figure and in Fig.~\ref{f3}(b) are
practically the same despite the fact that the minimal $T\tau_s$ value
in Fig.~\ref{f6} is six times less than that in Fig.~\ref{f3}(b)
($0.05$ against $0.3$). In this case, too, the experimental data are
well described by Eq.~(\ref{eq20}) which takes into account only the
Zeeman effect resulting in suppression of the two triplet channels in
growing magnetic field (see Fig.~\ref{f6}). The dashed curve in
Fig.~\ref{f6} is intended to illustrate once again the conflict between
our results and theoretical prediction for the role of SOI in the
interaction correction. It shows the temperature dependence of
$\delta\sigma^{ee}_{xx}$ for $B=0$, which would be observed if the
spin-orbit interaction would suppress the triplet contributions as
expected theoretically according to Eq.~(\ref{eq50}). One can see that
the correction calculated increases in magnitude with growing
temperature sufficiently steeper than the correction measured at
$B=0.2$~T so that the experimental slope of the $T$ dependence of
$\delta\sigma_{xx}^{ee}$ at $B\to 0$ is approximately four times less
than that found from Eq.~(\ref{eq50}) [shown in Fig.~\ref{f5} by the
circle and dotted line, respectively].

Thus, the  main experimental result of the paper is that the
suppression of the triplet contributions to interaction quantum
corrections by the spin-orbit interaction at $B=0$ does not reveal
itself under the experimental conditions when the spin relaxation rate
$1/\tau_s$ is $20-25$ times larger than $k_BT/\hbar$. All the
experimental results are consistently described within the framework of
the model, which takes into account the Zeeman effect only.

We would like to complete this work with a short discussion of the
possible reason for contradictions between the data and theoretical
expectations. Starting point of the theoretical considerations is that
the splitting of the energy spectrum due to the spin-orbit interaction
is supposed to be much smaller than all the characteristic energies in
the system. This approximation works good in the high-mobility systems,
in which the fast spin relaxation is accounted for by the large value
of the momentum relaxation time. It is not the case for the samples
under investigation. The value of the spin-orbit splitting estimated as
$\hbar\Omega=\hbar/\sqrt{2\tau\tau_s}$ is $(0.5-0.8)$~meV for different
electron densities, that is $10-15$ times larger than the lowest
temperature in our experiments, $T=0.6$~K. Furthermore, the spin-orbit
splitting, being significantly less than $\hbar/\tau$ for the rightmost
point in Fig.~\ref{f5} for which $\Omega\tau\simeq 0.15$, becomes close
to $\hbar/\tau$ for the leftmost point, where $\Omega\tau\simeq 0.8$
due to simultaneous increase of $\Omega$ and $\tau$ with increasing
electron density. Thus, the closer the value of $\Omega\tau$ to unity,
the stronger the disagreement between the data and theory. Quite
apparently, the spin-orbit splitting of the energy spectrum and
multicomponent character of the wave functions should be taken into
account from the outset when the interaction-induced correction is
estimated theoretically for the systems with strong spin-orbit
interaction.

\section{Conclusion}

We have studied the electron-electron interaction correction to the
conductivity of 2D electron gas in the gated single quantum well
InP/In$_{0.53}$Ga$_{0.47}$As heterostructures with strong spin-orbit
interaction. Analyzing the low magnetic field magnetoconductivity we
have obtained the value of the spin relaxation rate for different
electron density values, which appears to be relatively high as
compared with the temperature. The diffusion part of the interaction
correction has been obtained with the use of method based on the unique
property of the interaction to contribute to $\sigma_{xx}$ and do not
to $\sigma_{xy}$. It has been found that the behavior of the
interaction correction with the changing temperature and magnetic field
is analogous to that observed in the systems with slow spin relaxation.
Although the strong inequality $1/T\tau_s=20-25\gg 1$ is fulfilled
under our experimental conditions, both the temperature and magnetic
field dependences of the interaction correction are quantitatively
described in framework of the model developed for $1/T\tau_s\ll 1$, in
which  the Zeeman effect rather than the spin relaxation due to
spin-orbit coupling plays the main role.

Thus, despite the fast spin relaxation in the samples under
investigation, we do not observe any suppression of the triplet
contributions to the interaction quantum corrections to the
conductivity of 2D electron gas in the InP/In$_x$Ga$_{1-x}$As/InP
single quantum wells.

\section*{Acknowledgments}

We are grateful to I.~Gornyi and I.~Burmistrov for very useful
discussions. We would like to thank P.~Pool and S.~Studenikin for
providing with a sample with optimal mobility for this experiment. The
authors acknowledge the assistance by M.~Glasov in numerical
calculations. This work has been supported in part by the RFBR (Grant
Nos. 09-02-12206, 09-02-00789, 10-02-91336, and 10-02-00481).


\begin{thebibliography}{42}%
\makeatletter
\providecommand \@ifxundefined [1]{%
 \@ifx{#1\undefined}
}%
\providecommand \@ifnum [1]{%
 \ifnum #1\expandafter \@firstoftwo
 \else \expandafter \@secondoftwo
 \fi
}%
\providecommand \@ifx [1]{%
 \ifx #1\expandafter \@firstoftwo
 \else \expandafter \@secondoftwo
 \fi
}%
\providecommand \natexlab [1]{#1}%
\providecommand \enquote  [1]{``#1''}%
\providecommand \bibnamefont  [1]{#1}%
\providecommand \bibfnamefont [1]{#1}%
\providecommand \citenamefont [1]{#1}%
\providecommand \href@noop [0]{\@secondoftwo}%
\providecommand \href [0]{\begingroup \@sanitize@url \@href}%
\providecommand \@href[1]{\@@startlink{#1}\@@href}%
\providecommand \@@href[1]{\endgroup#1\@@endlink}%
\providecommand \@sanitize@url [0]{\catcode `\\12\catcode `\$12\catcode
  `\&12\catcode `\#12\catcode `\^12\catcode `\_12\catcode `\%12\relax}%
\providecommand \@@startlink[1]{}%
\providecommand \@@endlink[0]{}%
\providecommand \url  [0]{\begingroup\@sanitize@url \@url }%
\providecommand \@url [1]{\endgroup\@href {#1}{\urlprefix }}%
\providecommand \urlprefix  [0]{URL }%
\providecommand \Eprint [0]{\href }%
\@ifxundefined \urlstyle {%
  \providecommand \doi  [0]{\begingroup \@sanitize@url \@doi}%
  \providecommand \@doi [1]{\endgroup \@@startlink {\doibase
  #1}doi:\discretionary {}{}{}#1\@@endlink }%
}{%
  \providecommand \doi  [0]{doi:\discretionary{}{}{}\begingroup
  \urlstyle{rm}\Url }%
}%
\providecommand \doibase [0]{http://dx.doi.org/}%
\providecommand \Doi [0]{\begingroup \@sanitize@url \@Doi }%
\providecommand \@Doi  [1]{\endgroup\@@startlink{\doibase#1}\@@Doi}%
\providecommand \@@Doi [1]{#1\@@endlink}%
\providecommand \selectlanguage [0]{\@gobble}%
\providecommand \bibinfo  [0]{\@secondoftwo}%
\providecommand \bibfield  [0]{\@secondoftwo}%
\providecommand \translation [1]{[#1]}%
\providecommand \BibitemOpen [0]{}%
\providecommand \bibitemStop [0]{}%
\providecommand \bibitemNoStop [0]{.\EOS\space}%
\providecommand \EOS [0]{\spacefactor3000\relax}%
\providecommand \BibitemShut  [1]{\csname bibitem#1\endcsname}%
\bibitem [{\citenamefont {Hikami}\ \emph {et~al.}(1980)\citenamefont
    {Hikami},
  \citenamefont {Larkin},\ and\ \citenamefont {Nagaoka}}]{Hik80}%
  \BibitemOpen
  \bibfield  {author} {\bibinfo {author} {\bibfnamefont {S.}~\bibnamefont
  {Hikami}}, \bibinfo {author} {\bibfnamefont {A.~I.}\ \bibnamefont {Larkin}},
  \ and\ \bibinfo {author} {\bibfnamefont {Y.}~\bibnamefont {Nagaoka}},\
  }\href@noop {} {\bibfield  {journal} {\bibinfo  {journal} {Prog. Theor.
  Phys.},\ }\textbf {\bibinfo {volume} {63}},\ \bibinfo {pages} {707 }
  (\bibinfo {year} {1980})}\BibitemShut {NoStop}%
\bibitem [{\citenamefont {Chakravarty}\ and\ \citenamefont
  {Schmid}(1986)}]{Chak86}%
  \BibitemOpen
  \bibfield  {author} {\bibinfo {author} {\bibfnamefont {S.}~\bibnamefont
  {Chakravarty}}\ and\ \bibinfo {author} {\bibfnamefont {A.}~\bibnamefont
  {Schmid}},\ }\href@noop {} {\bibfield  {journal} {\bibinfo  {journal} {Phys.
  Reports},\ }\textbf {\bibinfo {volume} {140}},\ \bibinfo {pages} {193 }
  (\bibinfo {year} {1986})}\BibitemShut {NoStop}%
\bibitem [{\citenamefont {Wittmann}\ and\ \citenamefont
  {Schmid}(1987)}]{Wit87}%
  \BibitemOpen
  \bibfield  {author} {\bibinfo {author} {\bibfnamefont {H.-P.}\ \bibnamefont
  {Wittmann}}\ and\ \bibinfo {author} {\bibfnamefont {A.}~\bibnamefont
  {Schmid}},\ }\href@noop {} {\bibfield  {journal} {\bibinfo  {journal} {J. Low
  Temp. Phys.},\ }\textbf {\bibinfo {volume} {69}},\ \bibinfo {pages} {131 }
  (\bibinfo {year} {1987})}\BibitemShut {NoStop}%
\bibitem [{\citenamefont {Dresselhaus}\ \emph
    {et~al.}(1992)\citenamefont
  {Dresselhaus}, \citenamefont {Papavassiliou}, \citenamefont {Wheeler},\ and\
  \citenamefont {Sacks}}]{Dres92}%
  \BibitemOpen
  \bibfield  {author} {\bibinfo {author} {\bibfnamefont {P.~D.}\ \bibnamefont
  {Dresselhaus}}, \bibinfo {author} {\bibfnamefont {C.~M.~M.}\ \bibnamefont
  {Papavassiliou}}, \bibinfo {author} {\bibfnamefont {R.~G.}\ \bibnamefont
  {Wheeler}}, \ and\ \bibinfo {author} {\bibfnamefont {R.~N.}\ \bibnamefont
  {Sacks}},\ }\href@noop {} {\bibfield  {journal} {\bibinfo  {journal} {Phys.
  Rev. Lett.},\ }\textbf {\bibinfo {volume} {68}},\ \bibinfo {pages} {106 }
  (\bibinfo {year} {1992})}\BibitemShut {NoStop}%
\bibitem [{\citenamefont {Chen}\ \emph {et~al.}(1993)\citenamefont
    {Chen},
  \citenamefont {Han}, \citenamefont {Huang}, \citenamefont {Datta},\ and\
  \citenamefont {Janes}}]{Chen93}%
  \BibitemOpen
  \bibfield  {author} {\bibinfo {author} {\bibfnamefont {G.~L.}\ \bibnamefont
  {Chen}}, \bibinfo {author} {\bibfnamefont {J.}~\bibnamefont {Han}}, \bibinfo
  {author} {\bibfnamefont {T.~T.}\ \bibnamefont {Huang}}, \bibinfo {author}
  {\bibfnamefont {S.}~\bibnamefont {Datta}}, \ and\ \bibinfo {author}
  {\bibfnamefont {D.~B.}\ \bibnamefont {Janes}},\ }\href@noop {} {\bibfield
  {journal} {\bibinfo  {journal} {Phys. Rev. B},\ }\textbf {\bibinfo {volume}
  {47}},\ \bibinfo {pages} {4084 } (\bibinfo {year} {1993})}\BibitemShut
  {NoStop}%
\bibitem [{\citenamefont {Knap}\ \emph
  {et~al.}(1996){\natexlab{a}}\citenamefont {Knap}, \citenamefont {Zduniak},
  \citenamefont {Dmowski}, \citenamefont {Contreras},\ and\ \citenamefont
  {Dyakonov.}}]{Knap96}%
  \BibitemOpen
  \bibfield  {author} {\bibinfo {author} {\bibfnamefont {W.}~\bibnamefont
  {Knap}}, \bibinfo {author} {\bibfnamefont {A.}~\bibnamefont {Zduniak}},
  \bibinfo {author} {\bibfnamefont {L.~H.}\ \bibnamefont {Dmowski}}, \bibinfo
  {author} {\bibfnamefont {S.}~\bibnamefont {Contreras}}, \ and\ \bibinfo
  {author} {\bibfnamefont {M.~I.}\ \bibnamefont {Dyakonov.}},\ }\href@noop {}
  {\bibfield  {journal} {\bibinfo  {journal} {Phys. Stat. Sol. (b)},\ }\textbf
  {\bibinfo {volume} {198}},\ \bibinfo {pages} {267 } (\bibinfo {year}
  {1996}{\natexlab{a}})}\BibitemShut {NoStop}%
\bibitem [{\citenamefont {Knap}\ \emph
  {et~al.}(1996){\natexlab{b}}\citenamefont {Knap}, \citenamefont
  {Skierbiszewski}, \citenamefont {Zduniak}, \citenamefont {Litwin-Staszewska},
  \citenamefont {Bertho}, \citenamefont {Kobbi}, \citenamefont {Robert},
  \citenamefont {Pikus}, \citenamefont {Pikus}, \citenamefont {Iordanskii},
  \citenamefont {Mosser}, \citenamefont {Zekentes},\ and\ \citenamefont
  {Lyanda-Geller.}}]{Knap96-1}%
  \BibitemOpen
  \bibfield  {author} {\bibinfo {author} {\bibfnamefont {W.}~\bibnamefont
  {Knap}}, \bibinfo {author} {\bibfnamefont {C.}~\bibnamefont
  {Skierbiszewski}}, \bibinfo {author} {\bibfnamefont {A.}~\bibnamefont
  {Zduniak}}, \bibinfo {author} {\bibfnamefont {E.}~\bibnamefont
  {Litwin-Staszewska}}, \bibinfo {author} {\bibfnamefont {D.}~\bibnamefont
  {Bertho}}, \bibinfo {author} {\bibfnamefont {F.}~\bibnamefont {Kobbi}},
  \bibinfo {author} {\bibfnamefont {J.~L.}\ \bibnamefont {Robert}}, \bibinfo
  {author} {\bibfnamefont {G.~E.}\ \bibnamefont {Pikus}}, \bibinfo {author}
  {\bibfnamefont {F.~G.}\ \bibnamefont {Pikus}}, \bibinfo {author}
  {\bibfnamefont {S.~V.}\ \bibnamefont {Iordanskii}}, \bibinfo {author}
  {\bibfnamefont {V.}~\bibnamefont {Mosser}}, \bibinfo {author} {\bibfnamefont
  {K.}~\bibnamefont {Zekentes}}, \ and\ \bibinfo {author} {\bibfnamefont
  {Y.~B.}\ \bibnamefont {Lyanda-Geller.}},\ }\href@noop {} {\bibfield
  {journal} {\bibinfo  {journal} {Phys. Rev. B},\ }\textbf {\bibinfo {volume}
  {53}},\ \bibinfo {pages} {3912 } (\bibinfo {year}
  {1996}{\natexlab{b}})}\BibitemShut {NoStop}%
\bibitem [{\citenamefont {Lyanda-Geller}(1998)}]{Lyanda98}%
  \BibitemOpen
  \bibfield  {author} {\bibinfo {author} {\bibfnamefont {Y.}~\bibnamefont
  {Lyanda-Geller}},\ }\Doi {10.1103/PhysRevLett.80.4273} {\bibfield  {journal}
  {\bibinfo  {journal} {Phys. Rev. Lett.},\ }\textbf {\bibinfo {volume} {80}},\
  \bibinfo {pages} {4273} (\bibinfo {year} {1998})}\BibitemShut {NoStop}%
\bibitem [{\citenamefont {Altshuler}\ and\ \citenamefont
  {Aronov}(1985)}]{AA85}%
  \BibitemOpen
  \bibfield  {author} {\bibinfo {author} {\bibfnamefont {B.~L.}\ \bibnamefont
  {Altshuler}}\ and\ \bibinfo {author} {\bibfnamefont {A.~G.}\ \bibnamefont
  {Aronov}},\ }in\ \href@noop {} {\emph {\bibinfo {booktitle}
  {Electron-Electron Interaction in Disordered Systems}}},\ \bibinfo {editor}
  {edited by\ \bibinfo {editor} {\bibfnamefont {A.~L.}\ \bibnamefont {Efros}}\
  and\ \bibinfo {editor} {\bibfnamefont {M.}~\bibnamefont {Pollak}}}\ (\bibinfo
   {publisher} {North Holland},\ \bibinfo {address} {Amsterdam},\ \bibinfo
  {year} {1985})\ p.~\bibinfo {pages} {1}\BibitemShut {NoStop}%
\bibitem [{\citenamefont {Zala}\ \emph {et~al.}(2001)\citenamefont
    {Zala},
  \citenamefont {Narozhny},\ and\ \citenamefont {Aleiner}}]{Zala01}%
  \BibitemOpen
  \bibfield  {author} {\bibinfo {author} {\bibfnamefont {G.}~\bibnamefont
  {Zala}}, \bibinfo {author} {\bibfnamefont {B.~N.}\ \bibnamefont {Narozhny}},
  \ and\ \bibinfo {author} {\bibfnamefont {I.~L.}\ \bibnamefont {Aleiner}},\
  }\Doi {10.1103/PhysRevB.64.214204} {\bibfield  {journal} {\bibinfo  {journal}
  {Phys. Rev. B},\ }\textbf {\bibinfo {volume} {64}},\ \bibinfo {pages}
  {214204} (\bibinfo {year} {2001})}\BibitemShut {NoStop}%
\bibitem [{\citenamefont {Castellani}\ \emph
  {et~al.}(1984){\natexlab{a}}\citenamefont {Castellani}, \citenamefont
  {Di~Castro}, \citenamefont {Lee}, \citenamefont {Ma}, \citenamefont
  {Sorella},\ and\ \citenamefont {Tabet}}]{Cast84-2}%
  \BibitemOpen
  \bibfield  {author} {\bibinfo {author} {\bibfnamefont {C.}~\bibnamefont
  {Castellani}}, \bibinfo {author} {\bibfnamefont {C.}~\bibnamefont
  {Di~Castro}}, \bibinfo {author} {\bibfnamefont {P.~A.}\ \bibnamefont {Lee}},
  \bibinfo {author} {\bibfnamefont {M.}~\bibnamefont {Ma}}, \bibinfo {author}
  {\bibfnamefont {S.}~\bibnamefont {Sorella}}, \ and\ \bibinfo {author}
  {\bibfnamefont {E.}~\bibnamefont {Tabet}},\ }\Doi {10.1103/PhysRevB.30.1596}
  {\bibfield  {journal} {\bibinfo  {journal} {Phys. Rev. B},\ }\textbf
  {\bibinfo {volume} {30}},\ \bibinfo {pages} {1596} (\bibinfo {year}
  {1984}{\natexlab{a}})}\BibitemShut {NoStop}%
\bibitem [{\citenamefont {Finkel'stein}(1984){\natexlab{a}}}]{Fin84}%
  \BibitemOpen
  \bibfield  {author} {\bibinfo {author} {\bibfnamefont {A.~M.}\ \bibnamefont
  {Finkel'stein}},\ }\href@noop {} {\bibfield  {journal} {\bibinfo  {journal}
  {Zh. Eksp. Teor. Fiz.},\ }\textbf {\bibinfo {volume} {86}},\ \bibinfo {pages}
  {367} (\bibinfo {year} {1984}{\natexlab{a}})},\ \translation{Sov. Phys. JETP
  \textbf{59}, 212 (1984)}\BibitemShut {NoStop}%
\bibitem [{\citenamefont {Raimondi}\ \emph {et~al.}(1990)\citenamefont
  {Raimondi}, \citenamefont {Castellani},\ and\ \citenamefont
  {Di~Castro}}]{Raim90}%
  \BibitemOpen
  \bibfield  {author} {\bibinfo {author} {\bibfnamefont {R.}~\bibnamefont
  {Raimondi}}, \bibinfo {author} {\bibfnamefont {C.}~\bibnamefont
  {Castellani}}, \ and\ \bibinfo {author} {\bibfnamefont {C.}~\bibnamefont
  {Di~Castro}},\ }\Doi {10.1103/PhysRevB.42.4724} {\bibfield  {journal}
  {\bibinfo  {journal} {Phys. Rev. B},\ }\textbf {\bibinfo {volume} {42}},\
  \bibinfo {pages} {4724} (\bibinfo {year} {1990})}\BibitemShut {NoStop}%
\bibitem [{\citenamefont {Castellani}\ \emph
    {et~al.}(1998)\citenamefont
  {Castellani}, \citenamefont {Di~Castro},\ and\ \citenamefont {Lee}}]{Cast98}%
  \BibitemOpen
  \bibfield  {author} {\bibinfo {author} {\bibfnamefont {C.}~\bibnamefont
  {Castellani}}, \bibinfo {author} {\bibfnamefont {C.}~\bibnamefont
  {Di~Castro}}, \ and\ \bibinfo {author} {\bibfnamefont {P.~A.}\ \bibnamefont
  {Lee}},\ }\Doi {10.1103/PhysRevB.57.R9381} {\bibfield  {journal} {\bibinfo
  {journal} {Phys. Rev. B},\ }\textbf {\bibinfo {volume} {57}},\ \bibinfo
  {pages} {R9381} (\bibinfo {year} {1998})}\BibitemShut {NoStop}%
\bibitem [{\citenamefont {Zala}\ \emph {et~al.}(2002)\citenamefont
    {Zala},
  \citenamefont {Narozhny},\ and\ \citenamefont {Aleiner.}}]{Zala02}%
  \BibitemOpen
  \bibfield  {author} {\bibinfo {author} {\bibfnamefont {G.}~\bibnamefont
  {Zala}}, \bibinfo {author} {\bibfnamefont {B.~N.}\ \bibnamefont {Narozhny}},
  \ and\ \bibinfo {author} {\bibfnamefont {I.~L.}\ \bibnamefont {Aleiner.}},\
  }\href@noop {} {\bibfield  {journal} {\bibinfo  {journal} {Phys. Rev. B},\
  }\textbf {\bibinfo {volume} {65}},\ \bibinfo {pages} {020201} (\bibinfo
  {year} {2002})}\BibitemShut {NoStop}%
\bibitem [{\citenamefont {Gornyi}\ and\ \citenamefont {Mirlin}(2004)}]{Gor04}%
  \BibitemOpen
  \bibfield  {author} {\bibinfo {author} {\bibfnamefont {I.~V.}\ \bibnamefont
  {Gornyi}}\ and\ \bibinfo {author} {\bibfnamefont {A.~D.}\ \bibnamefont
  {Mirlin}},\ }\Doi {10.1103/PhysRevB.69.045313} {\bibfield  {journal}
  {\bibinfo  {journal} {Phys. Rev. B},\ }\textbf {\bibinfo {volume} {69}},\
  \bibinfo {pages} {045313} (\bibinfo {year} {2004})}\BibitemShut {NoStop}%
\bibitem [{\citenamefont {Minkov}\ \emph {et~al.}(2005)\citenamefont
    {Minkov},
  \citenamefont {Sherstobitov}, \citenamefont {Germanenko}, \citenamefont
  {Rut}, \citenamefont {Larionova},\ and\ \citenamefont {Zvonkov}}]{Min05-1}%
  \BibitemOpen
  \bibfield  {author} {\bibinfo {author} {\bibfnamefont {G.~M.}\ \bibnamefont
  {Minkov}}, \bibinfo {author} {\bibfnamefont {A.~A.}\ \bibnamefont
  {Sherstobitov}}, \bibinfo {author} {\bibfnamefont {A.~V.}\ \bibnamefont
  {Germanenko}}, \bibinfo {author} {\bibfnamefont {O.~E.}\ \bibnamefont {Rut}},
  \bibinfo {author} {\bibfnamefont {V.~A.}\ \bibnamefont {Larionova}}, \ and\
  \bibinfo {author} {\bibfnamefont {B.~N.}\ \bibnamefont {Zvonkov}},\
  }\href@noop {} {\bibfield  {journal} {\bibinfo  {journal} {Phys. Rev. B},\
  }\textbf {\bibinfo {volume} {72}},\ \bibinfo {pages} {165325} (\bibinfo
  {year} {2005})}\BibitemShut {NoStop}%
\bibitem [{\citenamefont {Altshuler}\ and\ \citenamefont
  {Aronov}(1983)}]{Altshuler83}%
  \BibitemOpen
  \bibfield  {author} {\bibinfo {author} {\bibfnamefont {B.~L.}\ \bibnamefont
  {Altshuler}}\ and\ \bibinfo {author} {\bibfnamefont {A.~G.}\ \bibnamefont
  {Aronov}},\ }\href@noop {} {\bibfield  {journal} {\bibinfo  {journal} {Solid
  State Comm.},\ }\textbf {\bibinfo {volume} {46}},\ \bibinfo {pages} {429}
  (\bibinfo {year} {1983})}\BibitemShut {NoStop}%
\bibitem [{\citenamefont {Gornyi}\ \emph {et~al.}(1998)\citenamefont
    {Gornyi},
  \citenamefont {Dmitriev},\ and\ \citenamefont {Kachorovski}}]{Gornyi98}%
  \BibitemOpen
  \bibfield  {author} {\bibinfo {author} {\bibfnamefont {I.~V.}\ \bibnamefont
  {Gornyi}}, \bibinfo {author} {\bibfnamefont {A.~P.}\ \bibnamefont
  {Dmitriev}}, \ and\ \bibinfo {author} {\bibfnamefont {V.~Y.}\ \bibnamefont
  {Kachorovski}},\ }\href@noop {} {\bibfield  {journal} {\bibinfo  {journal}
  {Pis'ma Zh. Eksp. Teor. Fiz.},\ }\textbf {\bibinfo {volume} {68}},\ \bibinfo
  {pages} {314} (\bibinfo {year} {1998})},\ \translation{JETP Letters
  \textbf{68}, 338 (1998)}\BibitemShut {NoStop}%
\bibitem [{\citenamefont {Skvortsov}(1998)}]{Skvortsov98}%
  \BibitemOpen
  \bibfield  {author} {\bibinfo {author} {\bibfnamefont {M.~A.}\ \bibnamefont
  {Skvortsov}},\ }\href@noop {} {\bibfield  {journal} {\bibinfo  {journal}
  {Pis'ma Zh. Eksp. Teor. Fiz.},\ }\textbf {\bibinfo {volume} {67}},\ \bibinfo
  {pages} {118} (\bibinfo {year} {1998})},\ \translation{JETP Lett.
  \textbf{67}, 133 (1998).}\BibitemShut {Stop}%
\bibitem [{\citenamefont {Minkov}\ \emph {et~al.}(2003)\citenamefont
    {Minkov},
  \citenamefont {Rut}, \citenamefont {Germanenko}, \citenamefont
  {Sherstobitov}, \citenamefont {Shashkin}, \citenamefont {Khrykin},\ and\
  \citenamefont {Zvonkov}}]{Min03-2}%
  \BibitemOpen
  \bibfield  {author} {\bibinfo {author} {\bibfnamefont {G.~M.}\ \bibnamefont
  {Minkov}}, \bibinfo {author} {\bibfnamefont {O.~E.}\ \bibnamefont {Rut}},
  \bibinfo {author} {\bibfnamefont {A.~V.}\ \bibnamefont {Germanenko}},
  \bibinfo {author} {\bibfnamefont {A.~A.}\ \bibnamefont {Sherstobitov}},
  \bibinfo {author} {\bibfnamefont {V.~I.}\ \bibnamefont {Shashkin}}, \bibinfo
  {author} {\bibfnamefont {O.~I.}\ \bibnamefont {Khrykin}}, \ and\ \bibinfo
  {author} {\bibfnamefont {B.~N.}\ \bibnamefont {Zvonkov}},\ }\Doi
  {10.1103/PhysRevB.67.205306} {\bibfield  {journal} {\bibinfo  {journal}
  {Phys. Rev. B},\ }\textbf {\bibinfo {volume} {67}},\ \bibinfo {pages}
  {205306} (\bibinfo {year} {2003})}\BibitemShut {NoStop}%
\bibitem [{\citenamefont {Minkov}\ \emph {et~al.}(2006)\citenamefont
    {Minkov},
  \citenamefont {Germanenko}, \citenamefont {Rut}, \citenamefont
  {Sherstobitov}, \citenamefont {Larionova}, \citenamefont {Bakarov},\ and\
  \citenamefont {Zvonkov}}]{Minkov06}%
  \BibitemOpen
  \bibfield  {author} {\bibinfo {author} {\bibfnamefont {G.~M.}\ \bibnamefont
  {Minkov}}, \bibinfo {author} {\bibfnamefont {A.~V.}\ \bibnamefont
  {Germanenko}}, \bibinfo {author} {\bibfnamefont {O.~E.}\ \bibnamefont {Rut}},
  \bibinfo {author} {\bibfnamefont {A.~A.}\ \bibnamefont {Sherstobitov}},
  \bibinfo {author} {\bibfnamefont {V.~A.}\ \bibnamefont {Larionova}}, \bibinfo
  {author} {\bibfnamefont {A.~K.}\ \bibnamefont {Bakarov}}, \ and\ \bibinfo
  {author} {\bibfnamefont {B.~N.}\ \bibnamefont {Zvonkov}},\ }\Doi
  {10.1103/PhysRevB.74.045314} {\bibfield  {journal} {\bibinfo  {journal}
  {Phys. Rev. B},\ }\textbf {\bibinfo {volume} {74}},\ \bibinfo {pages}
  {045314} (\bibinfo {year} {2006})}\BibitemShut {NoStop}%
\bibitem [{\citenamefont {Minkov}\ \emph
  {et~al.}(2007){\natexlab{a}}\citenamefont {Minkov}, \citenamefont
  {Germanenko}, \citenamefont {Rut}, \citenamefont {Sherstobitov},\ and\
  \citenamefont {Zvonkov}}]{Minkov07-1}%
  \BibitemOpen
  \bibfield  {author} {\bibinfo {author} {\bibfnamefont {G.~M.}\ \bibnamefont
  {Minkov}}, \bibinfo {author} {\bibfnamefont {A.~V.}\ \bibnamefont
  {Germanenko}}, \bibinfo {author} {\bibfnamefont {O.~E.}\ \bibnamefont {Rut}},
  \bibinfo {author} {\bibfnamefont {A.~A.}\ \bibnamefont {Sherstobitov}}, \
  and\ \bibinfo {author} {\bibfnamefont {B.~N.}\ \bibnamefont {Zvonkov}},\
  }\href@noop {} {\bibfield  {journal} {\bibinfo  {journal} {Phys. Rev. B},\
  }\textbf {\bibinfo {volume} {75}},\ \bibinfo {pages} {235316} (\bibinfo
  {year} {2007}{\natexlab{a}})}\BibitemShut {NoStop}%
\bibitem [{\citenamefont {Minkov}\ \emph {et~al.}(2009)\citenamefont
    {Minkov},
  \citenamefont {Germanenko}, \citenamefont {Rut}, \citenamefont
  {Sherstobitov},\ and\ \citenamefont {Zvonkov}}]{Minkov09}%
  \BibitemOpen
  \bibfield  {author} {\bibinfo {author} {\bibfnamefont {G.~M.}\ \bibnamefont
  {Minkov}}, \bibinfo {author} {\bibfnamefont {A.~V.}\ \bibnamefont
  {Germanenko}}, \bibinfo {author} {\bibfnamefont {O.~E.}\ \bibnamefont {Rut}},
  \bibinfo {author} {\bibfnamefont {A.~A.}\ \bibnamefont {Sherstobitov}}, \
  and\ \bibinfo {author} {\bibfnamefont {B.~N.}\ \bibnamefont {Zvonkov}},\
  }\href@noop {} {\bibfield  {journal} {\bibinfo  {journal} {Phys. Rev. B},\
  }\textbf {\bibinfo {volume} {79}},\ \bibinfo {pages} {235335} (\bibinfo
  {year} {2009})}\BibitemShut {NoStop}%
\bibitem [{\citenamefont {Studenikin}\ \emph
    {et~al.}(2003)\citenamefont
  {Studenikin}, \citenamefont {Coleridge}, \citenamefont {Ahmed}, \citenamefont
  {Poole},\ and\ \citenamefont {Sachrajda}}]{Studenikin03}%
  \BibitemOpen
  \bibfield  {author} {\bibinfo {author} {\bibfnamefont {S.~A.}\ \bibnamefont
  {Studenikin}}, \bibinfo {author} {\bibfnamefont {P.~T.}\ \bibnamefont
  {Coleridge}}, \bibinfo {author} {\bibfnamefont {N.}~\bibnamefont {Ahmed}},
  \bibinfo {author} {\bibfnamefont {P.~J.}\ \bibnamefont {Poole}}, \ and\
  \bibinfo {author} {\bibfnamefont {A.}~\bibnamefont {Sachrajda}},\ }\Doi
  {10.1103/PhysRevB.68.035317} {\bibfield  {journal} {\bibinfo  {journal}
  {Phys. Rev. B},\ }\textbf {\bibinfo {volume} {68}},\ \bibinfo {pages}
  {035317} (\bibinfo {year} {2003})}\BibitemShut {NoStop}%
\bibitem [{\citenamefont {Yu}\ \emph {et~al.}(2008)\citenamefont {Yu},
  \citenamefont {Dai}, \citenamefont {Chu}, \citenamefont {Poole},\ and\
  \citenamefont {Studenikin}}]{Yu08}%
  \BibitemOpen
  \bibfield  {author} {\bibinfo {author} {\bibfnamefont {G.}~\bibnamefont
  {Yu}}, \bibinfo {author} {\bibfnamefont {N.}~\bibnamefont {Dai}}, \bibinfo
  {author} {\bibfnamefont {J.~H.}\ \bibnamefont {Chu}}, \bibinfo {author}
  {\bibfnamefont {P.~J.}\ \bibnamefont {Poole}}, \ and\ \bibinfo {author}
  {\bibfnamefont {S.~A.}\ \bibnamefont {Studenikin}},\ }\Doi
  {10.1103/PhysRevB.78.035304} {\bibfield  {journal} {\bibinfo  {journal}
  {Phys. Rev. B},\ }\textbf {\bibinfo {volume} {78}},\ \bibinfo {pages}
  {035304} (\bibinfo {year} {2008})}\BibitemShut {NoStop}%
\bibitem [{\citenamefont {Zduniak}\ \emph {et~al.}(1997)\citenamefont
  {Zduniak}, \citenamefont {Dyakonov},\ and\ \citenamefont {Knap.}}]{Zdu97}%
  \BibitemOpen
  \bibfield  {author} {\bibinfo {author} {\bibfnamefont {A.}~\bibnamefont
  {Zduniak}}, \bibinfo {author} {\bibfnamefont {M.~I.}\ \bibnamefont
  {Dyakonov}}, \ and\ \bibinfo {author} {\bibfnamefont {W.}~\bibnamefont
  {Knap.}},\ }\href@noop {} {\bibfield  {journal} {\bibinfo  {journal} {Phys.
  Rev. B},\ }\textbf {\bibinfo {volume} {56}},\ \bibinfo {pages} {1996 }
  (\bibinfo {year} {1997})}\BibitemShut {NoStop}%
\bibitem [{\citenamefont {Golub}(2005)}]{Golub05}%
  \BibitemOpen
  \bibfield  {author} {\bibinfo {author} {\bibfnamefont {L.~E.}\ \bibnamefont
  {Golub}},\ }\Doi {10.1103/PhysRevB.71.235310} {\bibfield  {journal} {\bibinfo
   {journal} {Phys. Rev. B},\ }\textbf {\bibinfo {volume} {71}},\ \bibinfo
  {pages} {235310} (\bibinfo {year} {2005})}\BibitemShut {NoStop}%
\bibitem [{\citenamefont {Glazov}\ and\ \citenamefont
  {Golub}(2006)}]{Glazov06}%
  \BibitemOpen
  \bibfield  {author} {\bibinfo {author} {\bibfnamefont {M.~M.}\ \bibnamefont
  {Glazov}}\ and\ \bibinfo {author} {\bibfnamefont {L.~E.}\ \bibnamefont
  {Golub}},\ }\href@noop {} {\bibfield  {journal} {\bibinfo  {journal} {Fiz.
  Tekh. Poluprovodn.},\ }\textbf {\bibinfo {volume} {40}},\ \bibinfo {pages}
  {1241} (\bibinfo {year} {2006})},\ \translation{Semiconductors \textbf{40},
  1209 (2006)}\BibitemShut {NoStop}%
\bibitem [{\citenamefont {Finkel'stein}(1983)}]{Finkelstein83}%
  \BibitemOpen
  \bibfield  {author} {\bibinfo {author} {\bibfnamefont {A.~M.}\ \bibnamefont
  {Finkel'stein}},\ }\href@noop {} {\bibfield  {journal} {\bibinfo  {journal}
  {Zh. Eksp. Teor. Fiz.},\ }\textbf {\bibinfo {volume} {84}},\ \bibinfo {pages}
  {168} (\bibinfo {year} {1983})},\ \translation{Sov. Phys. JETP \textbf{57},
  97 (1983)}\BibitemShut {NoStop}%
\bibitem [{\citenamefont {Finkel'stein}(1984){\natexlab{b}}}]{Finkelstein84}%
  \BibitemOpen
  \bibfield  {author} {\bibinfo {author} {\bibfnamefont {A.~M.}\ \bibnamefont
  {Finkel'stein}},\ }\href@noop {} {\bibfield  {journal} {\bibinfo  {journal}
  {Z. Phys. B: Condens. Matter},\ }\textbf {\bibinfo {volume} {56}},\ \bibinfo
  {pages} {189} (\bibinfo {year} {1984}{\natexlab{b}})}\BibitemShut {NoStop}%
\bibitem [{\citenamefont {Castellani}\ \emph
  {et~al.}(1984){\natexlab{b}}\citenamefont {Castellani}, \citenamefont
  {Di~Castro}, \citenamefont {Lee},\ and\ \citenamefont {Ma}}]{Cast84-1}%
  \BibitemOpen
  \bibfield  {author} {\bibinfo {author} {\bibfnamefont {C.}~\bibnamefont
  {Castellani}}, \bibinfo {author} {\bibfnamefont {C.}~\bibnamefont
  {Di~Castro}}, \bibinfo {author} {\bibfnamefont {P.~A.}\ \bibnamefont {Lee}},
  \ and\ \bibinfo {author} {\bibfnamefont {M.}~\bibnamefont {Ma}},\ }\Doi
  {10.1103/PhysRevB.30.527} {\bibfield  {journal} {\bibinfo  {journal} {Phys.
  Rev. B},\ }\textbf {\bibinfo {volume} {30}},\ \bibinfo {pages} {527}
  (\bibinfo {year} {1984}{\natexlab{b}})}\BibitemShut {NoStop}%
\bibitem [{\citenamefont {Lefebvre}\ \emph {et~al.}(2002)\citenamefont
  {Lefebvre}, \citenamefont {Poole}, \citenamefont {Fraser}, \citenamefont
  {Aers}, \citenamefont {Chithrani},\ and\ \citenamefont
  {Williams}}]{Lefebvre02}%
  \BibitemOpen
  \bibfield  {author} {\bibinfo {author} {\bibfnamefont {J.}~\bibnamefont
  {Lefebvre}}, \bibinfo {author} {\bibfnamefont {P.~J.}\ \bibnamefont {Poole}},
  \bibinfo {author} {\bibfnamefont {J.}~\bibnamefont {Fraser}}, \bibinfo
  {author} {\bibfnamefont {G.~C.}\ \bibnamefont {Aers}}, \bibinfo {author}
  {\bibfnamefont {D.}~\bibnamefont {Chithrani}}, \ and\ \bibinfo {author}
  {\bibfnamefont {R.~L.}\ \bibnamefont {Williams}},\ }\href@noop {} {\bibfield
  {journal} {\bibinfo  {journal} {J. Cryst. Growth},\ }\textbf {\bibinfo
  {volume} {234}},\ \bibinfo {pages} {391} (\bibinfo {year}
  {2002})}\BibitemShut {NoStop}%
\bibitem [{\citenamefont {Narozhny}\ \emph {et~al.}(2002)\citenamefont
  {Narozhny}, \citenamefont {Zala},\ and\ \citenamefont {Aleiner}}]{Nar02}%
  \BibitemOpen
  \bibfield  {author} {\bibinfo {author} {\bibfnamefont {B.~N.}\ \bibnamefont
  {Narozhny}}, \bibinfo {author} {\bibfnamefont {G.}~\bibnamefont {Zala}}, \
  and\ \bibinfo {author} {\bibfnamefont {I.~L.}\ \bibnamefont {Aleiner}},\
  }\Doi {10.1103/PhysRevB.65.180202} {\bibfield  {journal} {\bibinfo  {journal}
  {Phys. Rev. B},\ }\textbf {\bibinfo {volume} {65}},\ \bibinfo {pages}
  {180202} (\bibinfo {year} {2002})}\BibitemShut {NoStop}%
\bibitem [{\citenamefont {Altshuler}\ \emph {et~al.}(1982)\citenamefont
  {Altshuler}, \citenamefont {Aronov},\ and\ \citenamefont
  {Khmelnitzkii}}]{Altshuler82}%
  \BibitemOpen
  \bibfield  {author} {\bibinfo {author} {\bibfnamefont {B.~L.}\ \bibnamefont
  {Altshuler}}, \bibinfo {author} {\bibfnamefont {A.~G.}\ \bibnamefont
  {Aronov}}, \ and\ \bibinfo {author} {\bibfnamefont {D.~E.}\ \bibnamefont
  {Khmelnitzkii}},\ }\href@noop {} {\bibfield  {journal} {\bibinfo  {journal}
  {J. Phys. C},\ }\textbf {\bibinfo {volume} {15}},\ \bibinfo {pages} {7367}
  (\bibinfo {year} {1982})}\BibitemShut {NoStop}%
\bibitem [{\citenamefont {Dyakonov}\ and\ \citenamefont
  {Perel}(1971)}]{Dyakonov71e}%
  \BibitemOpen
  \bibfield  {author} {\bibinfo {author} {\bibfnamefont {M.~I.}\ \bibnamefont
  {Dyakonov}}\ and\ \bibinfo {author} {\bibfnamefont {V.~I.}\ \bibnamefont
  {Perel}},\ }\href@noop {} {\bibfield  {journal} {\bibinfo  {journal} {Zh.
  Eksp. Teor. Fiz.},\ }\textbf {\bibinfo {volume} {60}},\ \bibinfo {pages}
  {1954} (\bibinfo {year} {1971})},\ \translation{Sov. Phys. JETP \textbf{33},
  1053 (1971)}\BibitemShut {NoStop}%
\bibitem [{\citenamefont {Bychkov}\ and\ \citenamefont
  {Rashba.}(1984)}]{Rash84}%
  \BibitemOpen
  \bibfield  {author} {\bibinfo {author} {\bibfnamefont {Y.~A.}\ \bibnamefont
  {Bychkov}}\ and\ \bibinfo {author} {\bibfnamefont {E.~I.}\ \bibnamefont
  {Rashba.}},\ }\href@noop {} {\bibfield  {journal} {\bibinfo  {journal} {J.
  Phys. C: Solid State Phys},\ }\textbf {\bibinfo {volume} {17}},\ \bibinfo
  {pages} {6039 } (\bibinfo {year} {1984})}\BibitemShut {NoStop}%
\bibitem [{\citenamefont {Minkov}\ \emph
  {et~al.}(2007){\natexlab{b}}\citenamefont {Minkov}, \citenamefont
  {Germanenko}, \citenamefont {Rut}, \citenamefont {Sherstobitov},\ and\
  \citenamefont {Zvonkov}}]{Minkov07}%
  \BibitemOpen
  \bibfield  {author} {\bibinfo {author} {\bibfnamefont {G.~M.}\ \bibnamefont
  {Minkov}}, \bibinfo {author} {\bibfnamefont {A.~V.}\ \bibnamefont
  {Germanenko}}, \bibinfo {author} {\bibfnamefont {O.~E.}\ \bibnamefont {Rut}},
  \bibinfo {author} {\bibfnamefont {A.~A.}\ \bibnamefont {Sherstobitov}}, \
  and\ \bibinfo {author} {\bibfnamefont {B.~N.}\ \bibnamefont {Zvonkov}},\
  }\Doi {10.1103/PhysRevB.76.165314} {\bibfield  {journal} {\bibinfo  {journal}
  {Phys. Rev. B},\ }\textbf {\bibinfo {volume} {76}},\ \bibinfo {pages}
  {165314} (\bibinfo {year} {2007}{\natexlab{b}})}\BibitemShut {NoStop}%
\bibitem [{\citenamefont {Mensz}\ and\ \citenamefont
  {Wheeler.}(1987)}]{Mens87}%
  \BibitemOpen
  \bibfield  {author} {\bibinfo {author} {\bibfnamefont {P.~M.}\ \bibnamefont
  {Mensz}}\ and\ \bibinfo {author} {\bibfnamefont {R.~G.}\ \bibnamefont
  {Wheeler.}},\ }\href@noop {} {\bibfield  {journal} {\bibinfo  {journal}
  {Phys. Rev. B},\ }\textbf {\bibinfo {volume} {35}},\ \bibinfo {pages} {2844 }
  (\bibinfo {year} {1987})}\BibitemShut {NoStop}%
\bibitem [{\citenamefont {Mathur}\ and\ \citenamefont
  {Baranger}(2001)}]{Math01}%
  \BibitemOpen
  \bibfield  {author} {\bibinfo {author} {\bibfnamefont {H.}~\bibnamefont
  {Mathur}}\ and\ \bibinfo {author} {\bibfnamefont {H.~U.}\ \bibnamefont
  {Baranger}},\ }\href@noop {} {\bibfield  {journal} {\bibinfo  {journal}
  {Phys. Rev. B},\ }\textbf {\bibinfo {volume} {64}},\ \bibinfo {pages}
  {235325} (\bibinfo {year} {2001})}\BibitemShut {NoStop}%
\bibitem [{\citenamefont {Minkov}\ \emph {et~al.}(2004)\citenamefont
    {Minkov},
  \citenamefont {Rut}, \citenamefont {Germanenko}, \citenamefont
  {Sherstobitov}, \citenamefont {Zvonkov}, \citenamefont {Shashkin},
  \citenamefont {Khrykin},\ and\ \citenamefont {Filatov}}]{Min04}%
  \BibitemOpen
  \bibfield  {author} {\bibinfo {author} {\bibfnamefont {G.~M.}\ \bibnamefont
  {Minkov}}, \bibinfo {author} {\bibfnamefont {O.~E.}\ \bibnamefont {Rut}},
  \bibinfo {author} {\bibfnamefont {A.~V.}\ \bibnamefont {Germanenko}},
  \bibinfo {author} {\bibfnamefont {A.~A.}\ \bibnamefont {Sherstobitov}},
  \bibinfo {author} {\bibfnamefont {B.~N.}\ \bibnamefont {Zvonkov}}, \bibinfo
  {author} {\bibfnamefont {V.~I.}\ \bibnamefont {Shashkin}}, \bibinfo {author}
  {\bibfnamefont {O.~I.}\ \bibnamefont {Khrykin}}, \ and\ \bibinfo {author}
  {\bibfnamefont {D.~O.}\ \bibnamefont {Filatov}},\ }\href@noop {} {\bibfield
  {journal} {\bibinfo  {journal} {Phys. Rev. B},\ }\textbf {\bibinfo {volume}
  {70}},\ \bibinfo {pages} {035304} (\bibinfo {year} {2004})}\BibitemShut
  {NoStop}%
\bibitem [{\citenamefont {Gornyi}()}]{PrvGornyi}%
  \BibitemOpen
  \bibfield  {author} {\bibinfo {author} {\bibfnamefont {I.~V.}\ \bibnamefont
  {Gornyi}},\ }\href@noop {} {\bibinfo  {journal} {private
  communication}}\BibitemShut {NoStop}%
\end{thebibliography}
\end{document}